\documentclass[%
 reprint,
superscriptaddress,
 amsmath,amssymb,longbibliography,
 aps,pra
]{revtex4-1}

\usepackage{setspace}
\usepackage{graphicx}
\usepackage{bbold}
\usepackage{amsmath}
\usepackage{amsfonts}
\usepackage{amsthm}
\usepackage{float}
\usepackage[margin=1in]{geometry}
\usepackage{enumerate}
\usepackage{etoolbox}
\usepackage{multirow}
\usepackage{amssymb}
\usepackage{comment}
\usepackage{mathrsfs}
\usepackage{ gensymb }
\usepackage{braket}
\usepackage{siunitx}
\usepackage{dcolumn}
\usepackage{subcaption}
\usepackage{xcolor}
\definecolor{dred}{rgb}{.8,0.2,.2}

\begin{document}

\newcommand{\moire}[0]{moir\'e\ }

\title{Sub-micrometer resolution neutron tomography}

\author{B. Heacock}
\email{bjheacoc@ncsu.edu}
\affiliation{Department of Physics, North Carolina State University, Raleigh, NC 27695, USA}
\affiliation{Triangle Universities Nuclear Laboratory, Durham, North Carolina 27708, USA}
\author{D. Sarenac}
\affiliation{Institute for Quantum Computing, University of Waterloo,  Waterloo, ON, Canada, N2L3G1}
\affiliation{Department of Physics, University of Waterloo, Waterloo, ON, Canada, N2L3G1}
\author{D. G. Cory}
\affiliation{Institute for Quantum Computing, University of Waterloo,  Waterloo, ON, Canada, N2L3G1} 
\affiliation{Department of Chemistry, University of Waterloo, Waterloo, ON, Canada, N2L3G1}
\affiliation{Perimeter Institute for Theoretical Physics, Waterloo, ON, Canada, N2L2Y5}
\affiliation{Canadian Institute for Advanced Research, Toronto, Ontario, Canada, M5G 1Z8}
\author{M. G. Huber}
\affiliation{National Institute of Standards and Technology, Gaithersburg, Maryland 20899, USA}
\author{J. P. W. MacLean}
\affiliation{Institute for Quantum Computing, University of Waterloo,  Waterloo, ON, Canada, N2L3G1}
\affiliation{Department of Physics, University of Waterloo, Waterloo, ON, Canada, N2L3G1}
\author{H. Miao}
\affiliation{Biophysics and Biochemistry Center, National Heart, Lung and Blood Institute, National Institutes of Health, Bethesda, Maryland USA}
\author{H. Wen}
\affiliation{Biophysics and Biochemistry Center, National Heart, Lung and Blood Institute, National Institutes of Health, Bethesda, Maryland USA}
\author{D. A. Pushin}
\email{dmitry.pushin@uwaterloo.ca}
\affiliation{Institute for Quantum Computing, University of Waterloo,  Waterloo, ON, Canada, N2L3G1}
\affiliation{Department of Physics, University of Waterloo, Waterloo, ON, Canada, N2L3G1}

\begin{abstract}
We demonstrate a neutron tomography technique with sub-micrometer spatial resolution. Our method consists of measuring neutron diffraction spectra using a double crystal diffractometer as a function of sample rotation and then using a phase retrieval algorithm followed by tomographic reconstruction to generate a density map of the sample.  In this first demonstration, silicon phase-gratings are used as samples, the periodic structure of which allows the shape of the gratings to be imaged without the need of position sensitive detectors.  Topological features found in the  reconstructions also appear in scanning electron micrographs.  The reconstructions have a resolution of about 300 nm, which is over an order of magnitude smaller than the resolution of radiographic, phase contrast, differential phase contrast, and dark field neutron tomography methods. Further optimization of the underlying phase recovery and tomographic reconstruction algorithm is also considered.
\end{abstract}

\maketitle

\section{Introduction}  

Neutron imaging and scattering provide a unique probe for a wide variety of materials, motivating the construction of a growing number of neutron imaging and scattering user-facilities \cite{chen2016china, garoby2017progress}.  Cold neutrons have wavelengths similar to x-rays, but whereas x-rays interact strongly with high-Z atoms, neutrons tend to scatter off of materials with a high hydrogen content or other light nuclei.  While many imaging and scattering techniques, such as radiography and computed tomography (CT), are shared between x-rays and neutrons, the two radiation sources provide complementary information \cite{strobl2009advances} and have even recently been combined in a single apparatus \cite{lamanna2017neutron,chiang2017simultaneous}.

For typical applications, neutron imaging is sensitive to sample features larger than roughly 20~$\mu$m, while neutron scattering probes smaller structures without producing images of the sample.  The spatial resolution of neutron imaging is limited by the resolution of neutron position sensitive detectors (PSDs) ($\sim 20 \; \mu \mathrm{m} $), though a range of techniques can push neutron imaging resolution down to a few micrometers \cite{hussey2017neutron, harti2017sub, williams2012detection,strobl2009advances}.    In particular, neutron CT is a very useful form of neutron imaging and has been demonstrated with radiographic, phase contrast, differential phase contrast, and dark field signals \cite{allman2000imaging,strobl2004small,pushin2006reciprocal,pushin2007reciprocal,strobl2008neutron,strobl2009advances}. We use neutron scattering data with a phase retrieval algorithm to perform neutron CT with a spatial resolution of about 300~nm, more than an order of magnitude smaller than other neutron imaging techniques. Our method consists of measuring neutron diffraction spectra as a function of sample rotation and using a phase retrieval algorithm to recover the phase in  position-space.  The recovered phase as a function of sample rotation is then tomographically reconstructed, giving a two-dimensional density map of the sample.  Because the diffraction data  characterizes the sample in Fourier space, the resolution of the resulting image is not dependent on the achievable resolution of neutron PSDs.

The neutron diffraction spectra were measured using a double crystal diffractometer, similar to typical ultra-small-angle neutron scattering (USANS) techniques.  In this first demonstration we use silicon phase-gratings as a sample, the periodic structure of which need only stay constant over the neutron's coherence length ($\sim 35 \; \mu \mathrm{m}$ in this case), rendering this technique insensitive to many types of sample defects.  Here, we image gratings with periods of 2.4~$\mu$m over a beam size of 4.4~mm, but the neutron coherence length and momentum-space resolution of the double crystal diffractometer are such that the resulting real-space image size is the width of ten to twenty grating periods.  While our technique is insensitive to a low-density of grating defects, such as a vacancies or dislocations, the overall shape and period of the grating is assumed to be uniform over the entire beam.  If a PSD were used instead of the proportional counter used in this experiment, the periodic portions of the sample would need only extend to the pixel size, putting the sub-micrometer imaging of quasi-periodic samples firmly in reach.  In practice, the per pixel count rate in such a setup would set an effective minimum beam size.  

The phase retrieval and tomographic reconstruction techniques demonstrated here may be useful for a sophisticated USANS spectrometer, such as those described in \cite{barker2005design} or \cite{strobl2007new} for a wide variety of sample types.    Phase retrieval and tomography is also used for x-ray coherent diffractive imaging (CDI) \cite{shechtman2015phase,rodriguez2013oversampling,martin2012noise,burvall2011phase,langer2008quantitative}, and phase retrieval using the transport of intensity equation (TIE) method for intermediate-field applications is used by phase-contrast neutron imaging \cite{allman2000imaging,strobl2009advances}. The techniques described here are also directly applicable to other far-field scattering measurements, such as small-angle neutron scattering (SANS) \cite{glinka199830}, where the measured Fourier images are inherently two-dimensional, instead of one-dimensional, as is the case for USANS spectrometers.

\section{Experiment}

\begin{figure}
\centering
\includegraphics[width=.9\linewidth]{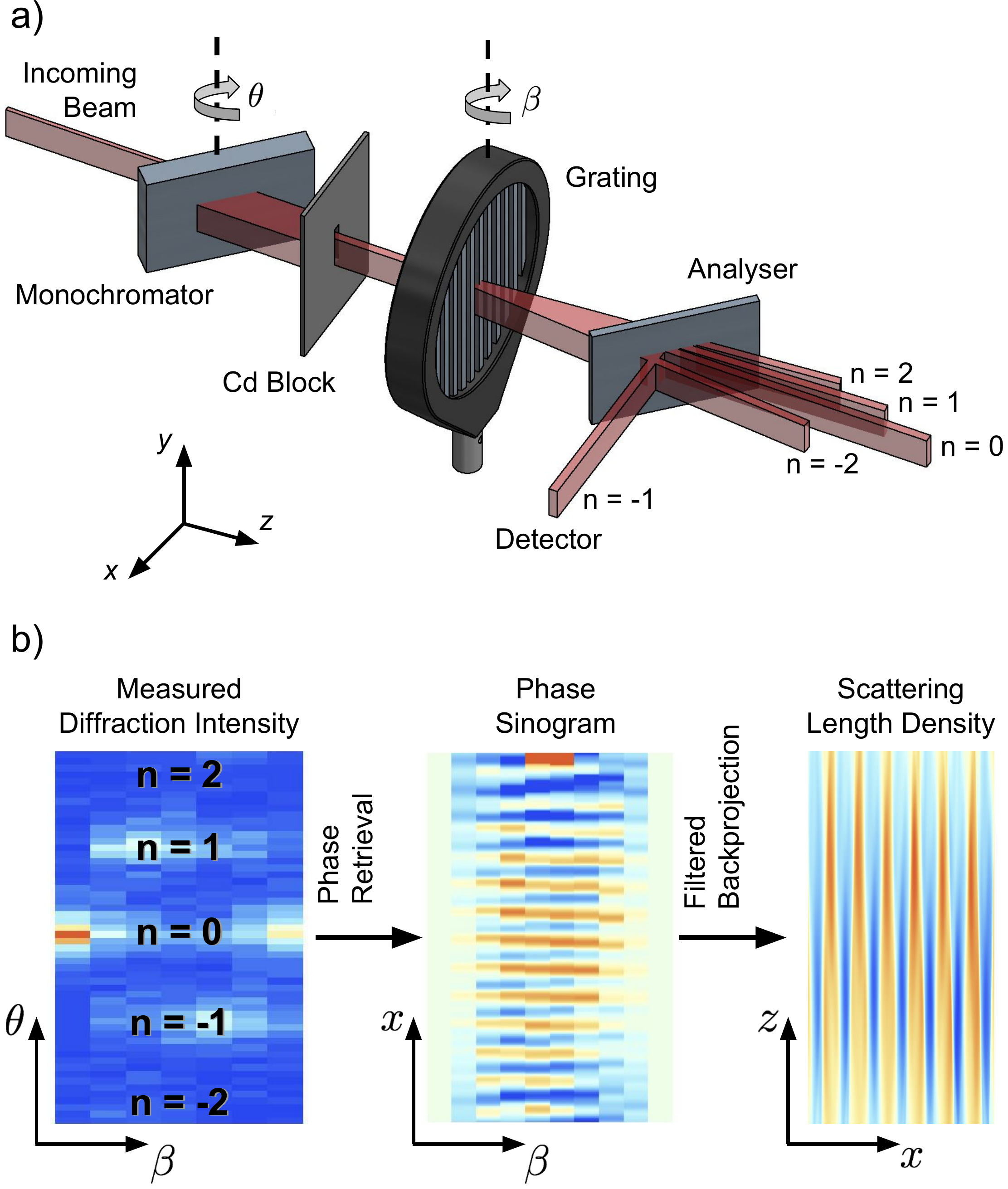}
\caption{(a) Experimental setup. A $\lambda=4.4$~\AA \; neutron beam passes through a monochromator crystal, then through a phase-grating whose effect is measured by an analyzer crystal and a $^3$He proportional counter.  (b) From the measured diffraction intensity, the position-space phase is retrieved, giving the phase sinogram, which is then used to reconstruct the scattering length density of the grating.}
 \label{fig:setup}
\end{figure}

The experiment  was performed at the NIOFa beamline at the National Institute of Standards and Technology (NIST) Center for Neutron Research (NCNR) in Gaithersburg, MD  \cite{shahi2016new,pushin2015neutron}. A schematic of the experiment is shown in Fig.~\ref{fig:setup}a. A 4.4~\AA \;wavelength neutron beam is extracted from the neutron guide using a pyrolytic graphite (PG) crystal. The beam passes through a 2~mm slit before being Bragg diffracted (Laue geometry) by a perfect-silicon crystal (111) monochromator. A 4.4~mm wide cadmium block is used to select the forward-diffracted beam from the monochromator. To measure the outgoing momentum distribution modified by the phase-grating, we placed a second perfect-silicon crystal (111) analyzer after the grating, forming a double crystal diffractometer.  The monochromator was rotated relative to the analyzer by a rotation stage with an embedded optical encoder, allowing arcsecond precision motion.

The Bragg diffracted wavepackets are Lorentzian in shape in momentum-space. Their nominal transverse coherence length is given by the pendell\"{o}sung length $\Delta_H = 35 \; \mu \mathrm{m}$ for the (111) reflection from silicon at $\lambda = 4.4$~\AA.\;  
Diffraction peaks with angular locations of

\begin{align}
\theta_n=\sin^{-1} \left(\frac{n\lambda}{\lambda_G}\right),
\label{Eqn:thetaquantum}
\end{align}

\noindent
where $\lambda_G$ is the period of the grating, and $n$ is an integer that represents the diffraction order, were clearly visible (Fig.~\ref{fig:setup}b).  The relative amplitudes of the diffraction peaks depend on the shape and amplitude of the phase profile for a given angle of sample rotation $\beta$.

\begin{figure}
\centering
\includegraphics[width=.98\linewidth]{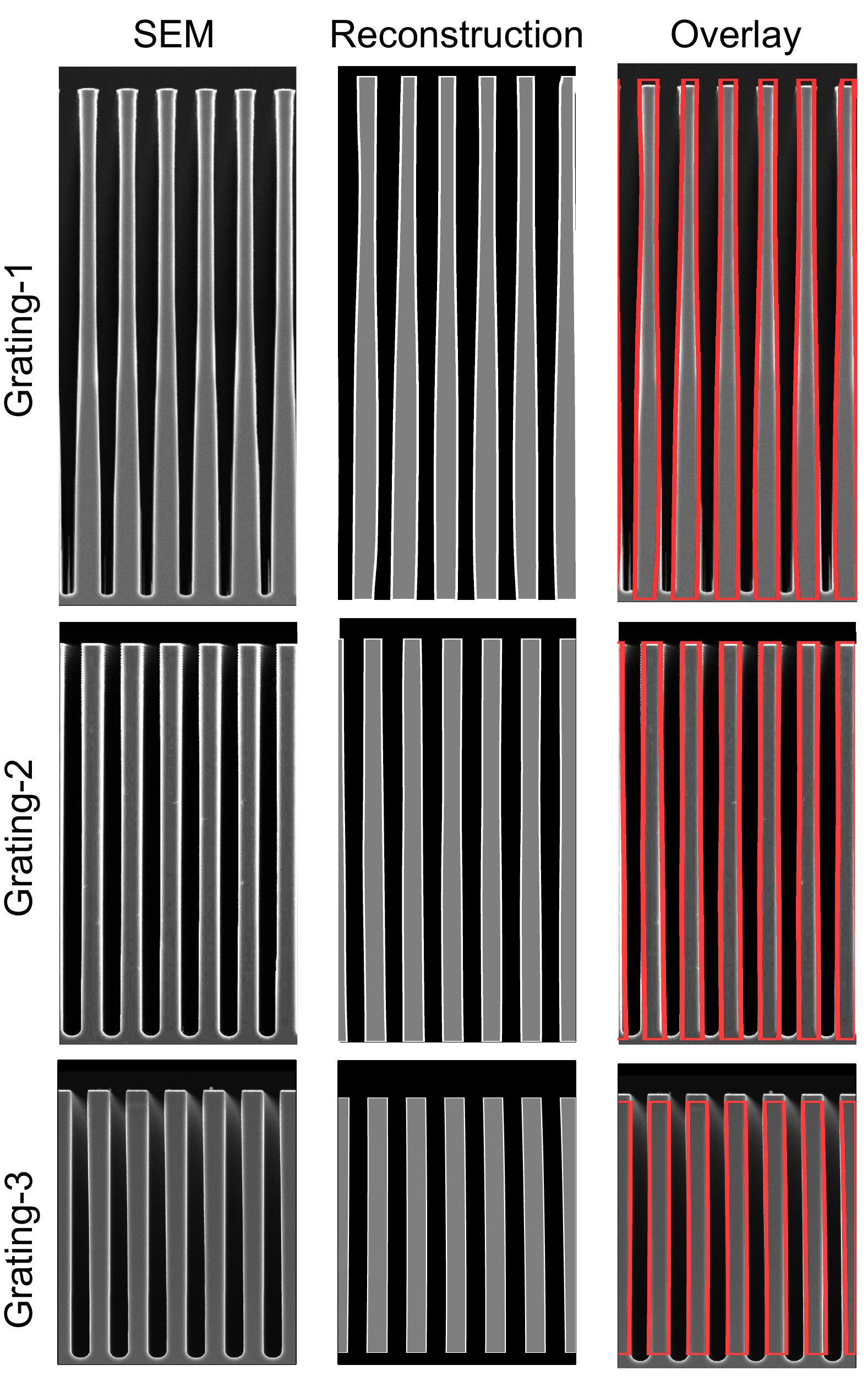}
\caption{SEM micrographs of the phase-gratings (left column) compared to their neutron tomographic reconstructions (middle column).  Also shown is an overlay (right column) of the outline of the reconstruction over the SEM.  The good agreement between the SEM and reconstructions indicate that the shape of the gratings are uniform over a large range.  The walls of Grating-2 and Grating-3 are shown to be very straight, while the sloped walls of Grating-1 appear in both the SEM micrograph and the reconstruction.  Edge highlights added to the reconstructions for clarity.}
 \label{fig:recon}
\end{figure}

Three gratings were analyzed in this experiment. The period of each grating was $\lambda_G=2.4$~$\mu$m. The grating depths were $h = 29.0~\mu$m, $h = 23.9~\mu$m, and $h = 15.8~\mu$m, with corresponding phase amplitudes of 2.6~rad, 2.2~rad, and 1.4~rad, respectively for $\lambda=4.4$~\AA\; neutrons. Scanning Electron Microscope (SEM) micrographs of the three gratings are shown in Fig.~\ref{fig:recon}.

Diffraction spectra as a function of grating rotation $\beta$ about the $y$-axis were taken from approximately -2~degrees to 4~degrees in 1 degree steps $y$-axis rotation for Grating-1, -6~degrees to 6~degrees in 1~degree steps for Grating-2, and -6~degrees to 6~degrees in 1.5~degree steps for Grating-3.  The -3~degree $y$-axis rotation diffraction spectrum for Grating-1 was substituted with the 5~degree spectrum, because the -3~degree spectrum was out of the measured range.  The grating rotation was severe enough for there to be no diffraction peaks expected in the -3~degree spectrum.

The position-space phase of each measured diffraction spectra was computed using a phase retrieval algorithm.  The retrieved phase as a function of $\beta$ forms a sinogram that was then tomographically reconstructed, giving two-dimensional images of the scattering length density of the gratings (Fig.~\ref{fig:setup}b).  See Section~\ref{sec:algorithm} for a detailed description of the reconstruction algorithm.  The reconstructions of the grating scattering length density and SEM micrographs of the three gratings are shown in Fig.~\ref{fig:recon}.  The vertical walls of Grating-2 and Grating-3 are captured by the reconstructions, while the slope and curvature of the walls of Grating-1 visible in the SEM micrograph are well-represented in the reconstruction.  The good agreement between the SEM micrographs and the reconstructions also imply that the grating profile is uniform over a much larger region than what is visible to the SEM.  A uniform phase profile is critical for neutron mori\'{e} interferometers, the recent demonstrations of which used the same phase-gratings \cite{pushin2017far,sarenac2018three}.

The spatial resolution of the reconstructions are a fraction of the 2.4~$\mu$m period.  We estimate the spatial resolution to be about one fourth of $\lambda_G / n_\mathrm{max}$.  The diffraction spectra were measured over a large enough range to resolve $n=2$ (which was highly suppressed for these phase-gratings), so the ultimate resolution is estimated to be 300~nm.

\begin{figure}
\centering
\includegraphics[width=.9\linewidth]{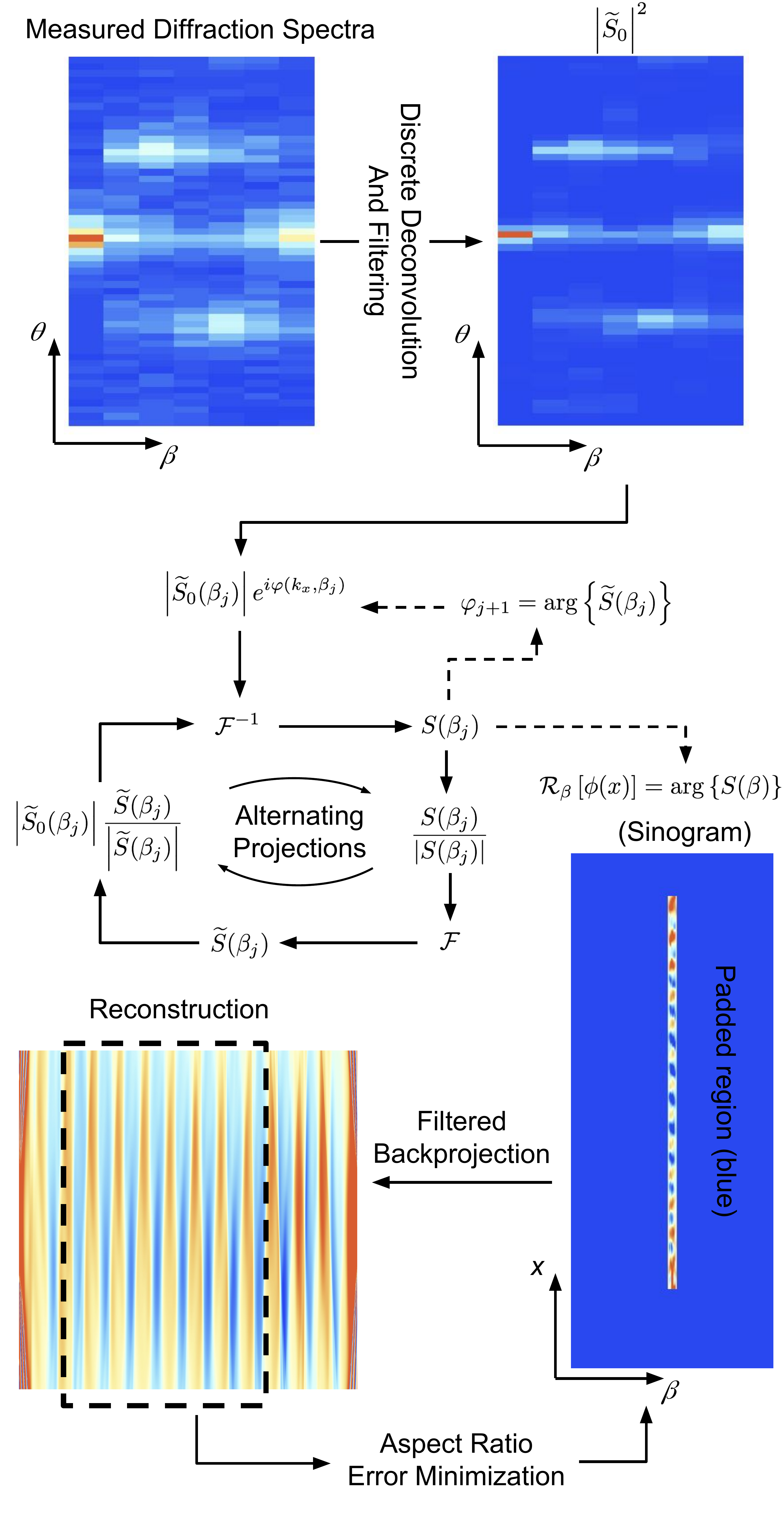}
\caption{Outline of the entire reconstruction process.  The raw data is filtered and deconvolved from the average diffraction spectrum of the maximum and minimum measured sample rotation angles.  The output is passed into an alternating projections algorithm for each sample rotation angle $\beta$, with the previous solution seeding the next step in $\beta$.  The retrieved position-space phase as a function of $x$ and $\beta$ forms a sinogram, which is padded with zeros so that $\beta$ extends from $-\pi/2$ to $\pi/2$.  A filtered back projection of the sinogram creates the reconstructed scattering length density.  The high-fidelity portion of the reconstruction is made into a binary image, Radon-transformed back into a sinogram, and compared to the original sinogram over the relevant range.  The sinogram error is minimized with respect to the aspect ratio of the reconstruction.  See text for details.}
 \label{fig:Alg}
\end{figure}

\section{Reconstruction Algorithm}
\label{sec:algorithm}

The phase profile $\phi(x)$ of a neutron propagating in the $z$-direction through a material is given in the Eikonal approximation by \cite{sears1989neutron}

\begin{equation}
\phi(x) = -\frac{1}{\hbar} \int dt \, V_0 \simeq - \lambda \int d z \, \left \langle b(x,z) \right \rangle,
\label{Eqn:Eikonal}
\end{equation}

\noindent
where $\lambda$ is the neutron wavelength; $V_0$ is the material's optical potential; the integral is taken over the neutron's trajectory; and $\langle b (x,z) \rangle $ is the spatially-dependent bound coherent scattering length density.  For example, in a homogeneous material $\langle b (x,z) \rangle = N b_c$ inside the material and $\langle b (x,z) \rangle = 0$ outside the material, where $N$ is the atomic number density and $b_c$ is the bound coherent scattering length.  The goal of tomography is then to reconstruct $\langle b(x,z) \rangle$.  We consider one-dimensional phase profiles, though the extension to two dimensions is straightforward, in which case the reconstruction of the scattering length density is three-dimensional instead of two-dimensional.

A sample that imprints a spatially-dependent phase $\phi(x)$ over the incoming wavefunction $\Psi_i(x)$, results in an outgoing wavefunction of

\begin{equation}
\Psi_{f}(x)=e^{-i \phi(x)}\Psi_{i}(x) .
\label{Eqn:psiout}
\end{equation}

To analyze sample diffraction we look at the neutron wavefunction in momentum-space: 

\begin{align}
\widetilde{\Psi}_{f}(k_x)=\mathcal{F}\{\Psi_{f}\}=\mathcal{F}\{e^{-i \phi(x)}\}*\mathcal{F}\{\Psi_{i}\} ,
\label{Eqn:psik}
\end{align}

\noindent 
where $\mathcal{F}\{...\}$ is a Fourier transform, and $*$ is the convolution operator. In this experiment, we measure the outgoing neutron momentum distribution, or diffraction spectrum, which is calculated as $P_f(k_x)=|\widetilde{\Psi}_f(k_x)|^2$.

If diffraction spectra are taken as a function of sample rotation, then the resulting momentum-space wave function is given by

\begin{equation}
\widetilde{\Psi}_f(k_x) = \widetilde{S}(k_x,\beta) * \widetilde{\Psi}_i,
\label{eqn:phiRot}
\end{equation}

\noindent
with the function $\widetilde{S}(k_x,\beta)$ defined in terms of the the grating rotation angle $\beta$ about the $y$-axis (see Fig.~\ref{fig:setup}) and 

\begin{equation}
\widetilde{S}(k_x,\beta) = \mathcal{F} \left \{  S(x,\beta) \right \},
\label{eqn:Sfour}
\end{equation}

\noindent
given that

\begin{equation}
S(x,\beta)= e^{- i \mathcal{R}_\beta \left [ \phi(x) \right ] },
\label{eqn:Sdef}
\end{equation}

\noindent
where integrating over the neutron's trajectory through a sample as a function of sample rotation modifies the effective phase profile (Eqn.~\ref{Eqn:Eikonal}) via a Radon transformation $\mathcal{R}_\beta [\phi(x) ]$.  For clarity, we drop the $k_x$ and $x$ arguments in $\widetilde{S}(\beta)$ and $S(\beta)$, respectively.  

Because the size of the beam is much larger than the grating period, when $P_f(k_x)$ is averaged over translations $x_0$ of the incoming state, $\widetilde{\Psi}_i(k_x) \rightarrow \widetilde{\Psi}_i(k_x)e^{i k_x x_0}$, the measured momentum distribution reduces to

\begin{equation}
P_{f}(k_x,\beta) = \left | \widetilde{S}(\beta) \right |^2 * \left |  \widetilde{\Psi}_i  \right |^2 .
\label{eqn:PfAvg}
\end{equation}

\noindent
This reduction can also be viewed as an averaging over the phase, or physical translation in the $x$-direction, of the periodic structure contained in $\phi(x)$.  Thus the result is independent of the translation of the periodic structure, rendering this technique insensitive to many types of sample defects. 

Before recovering the phase profile $\phi(x)$, the measured momentum distribution is deconvolved with $\left |\widetilde{\Psi}_i \right |^2$, which is simply the measured diffraction spectrum with no sample present.  For this experiment, discrete deconvolutions are performed between the measured momentum distributions for each $y$-axis rotation and the average of the first and last momentum distributions (largest grating rotations), where there were no visible diffraction peaks.  Additionally, a two-dimensional Gaussian filter is applied to the resulting $ \left | \widetilde{S} (\beta) \right |^2 $ with respect to diffraction angle $\theta$ and grating rotation $\beta$ for noise suppression.  While this step is sufficient for our purposes, there are other methods for modifying phase recovery algorithms for noisy data \cite{langer2008quantitative,martin2012noise,rodriguez2013oversampling}.  The deconvolved diffraction spectra $ \left | \widetilde{S}(\beta) \right |^2$ are thus isolated.  Then one can form

\begin{equation}
\mathcal{F}^{-1} \left \{ \left |  \widetilde{S}(\beta) \right | e^{i \varphi(k_x,\beta)} \right \} = e^{-i \mathcal{R}_\beta \left [ \phi (x) \right ] }
\label{eqn:phik}
\end{equation}

\noindent
where $\varphi(k_x,\beta)$ is an unknown function.  While the absolute value of  $\widetilde{S}(\beta)$ is known from the deconvolution of the incoming and outgoing momentum distributions, the phase $\varphi(k_x,\beta)$ has not been measured.  Both the momentum-space phase $\varphi(k_x,\beta)$ and the position-space phase $\mathcal{R}_\beta [ \phi (x) ]$ can be retrieved with an alternating projections algorithm \cite{shechtman2015phase}, an outline of which is shown in Fig.~\ref{fig:Alg}. 

\subsection{Phase Recovery}
\label{sec:phaserec}

Phase recovery by alternating projections works by alternating between real-space and Fourier-space magnitude constraints \cite{shechtman2015phase}.  In our case, the Fourier-space magnitude constraint step is imposed by updating $\widetilde{S}(\beta)$ according to

\begin{equation}
\widetilde{S}(\beta_j) \rightarrow \left | \widetilde{S}_0(\beta_j) \right | \frac{\widetilde{S}(\beta_j)}{
\left | \widetilde{S}(\beta_j)  \right |}
\label{eqn:kcon}
\end{equation}

\noindent
in Fig.~\ref{fig:Alg}, where $j$ indexes the grating rotation angle, and $ \left | S_0(\beta_j) \right |$ is the deconvolved measured diffraction spectrum.  The real-space magnitude constraint comes from assuming that absorbtion is negligible, in which case $| S(\beta)|^2 = 1$.  The function $S(\beta)$ is updated with each iteration of the algorithm by taking

\begin{equation}
S(\beta_j) \rightarrow \frac{S(\beta_j) } {\left | S(\beta_j)  \right | },
\label{eqn:xcon}
\end{equation}

\noindent
as shown in Fig.~\ref{fig:Alg}.  Note that there are methods for extending the real-space constraint or phase recovery in general to samples with non-neglible absorbtion \cite{shechtman2015phase, burvall2011phase}.  

It is well-known that alternating projection algorithms can be sensitive to the initial phase guess $\varphi(k_x,\beta)$, as many global minima are possible \cite{shechtman2015phase}.  Some of these minima may be physical, while others are not.  In general, solutions are impervious to complex conjugations, phase offsets, and real-space translations of $S(\beta)$ \cite{guizar2012understanding}.  This complicates phase retrieval for the purposes of tomography, because the solution space needs to be continuous between rotation steps, $\beta_j \rightarrow \beta_{j\pm1}$.  A number of phase retrieval algorithms for tomography exist \cite{burvall2011phase,langer2008quantitative}, but we find a suitable way to ensure a continuous solution space as a function of grating rotation is to initiate the next iteration of $\beta_{j\pm 1}$ with the previous solution of $\varphi(k_x,\beta_j)$.  Therefore continuity in $\beta$-space is imposed by taking

\begin{equation}
\varphi(k_x,\beta_{j \pm 1}) \rightarrow \varphi(k_x,\beta_j)
\label{eqn:betastep}
\end{equation}

\noindent
for an initial guess at each step in $\beta$.  In our algorithm, we do this from the middle out, choosing the initial value of $\beta$ to correspond to the grating approximately perpendicular with the beam, then seeding subsequent $\varphi(k_x,\beta)$ with that of the previous $\beta$ in both the positive and negative sample rotation directions.  

The initial $\widetilde{S}(\beta)$ is found by passing a random phase $\varphi(k_x,\beta)$ into the alternating projections algorithm.  This is repeated two hundred times, and the resulting solution with a minimum error in $\widetilde{S}(\beta)$ and $S(\beta)$ is selected.  The error is defined as

\begin{multline}
\epsilon(\beta) = \sum_{k_x}{ \left (\left|\widetilde{S}(\beta) \right |-\left |\widetilde{S}_0(\beta) \right | \right ) ^2} + \\
+ {1 \over N} \sum_x \Big ( \left | S ( \beta) \right| -1 \Big )^2  ,
\label{eqn:err}
\end{multline}

\noindent
where $N$ is the number of points in the $x$-dimension of the array representing $|S(\beta)|$.  This term gives equal weighting to the error in position and momentum-space, because the discrete deconvolution process normalizes $\widetilde{S}_0$, such that $\sum_{k_x}{ \left | \widetilde{S}_0(\beta) \right |^2}=1$

After the phase of $S(\beta)$ is successfully retrieved, a sinogram is formed with the region beyond maximum rotation padded with zeros.  The range of rotation that should be measured will depend on the aspect ratio of the periodic structure.  Once the rotation of the sample in radians exceeds its aspect ratio, the sample diffraction will be much weaker, if even measurable.  Thus padding the sinogram with zeros as described provides a good estimation of the full sinogram.

\subsection{Tomographic Reconstruction}

The tomographic reconstructions of the recovered position-space phase are completed with a filtered back projection \cite{strobl2009advances}.  In our case, we use a Hann filter, though other filtering functions may be selected.  The aspect ratio of the reconstructed image is found by performing the filtered back projection, forming a binary image, then Radon-transforming the resulting image and computing the error when compared to the original, recovered phase sinogram.  The error is minimized with respect to the aspect ratio of the reconstruction.  This is a local minimum in the neighborhood of the aspect ratio as predicted by the known scattering length density of silicon, the grating period measured using the diffraction peak positions and Eqn.~\ref{Eqn:thetaquantum}, and the amplitude of the recovered phase sinusoid when the grating is approximately aligned with the beam.  

The binarization of the reconstruction is completed by setting all the reconstruction values below zero to zero and all the values above zero to one.  The average value of the reconstructions is zero because of the previously-mentioned padding for large rotation angles.  The offset of the measured and padded portions of the phase sinogram does not affect the reconstruction because of the initial filtering step in the filtered back projection.  Note that other methods for discrete tomography \cite{krimmel2005discrete} may also be useful for samples made of a single material and cut to a certain shape. A comparison of the reconstructions and the SEM are shown in Fig.~\ref{fig:recon}. The grating color was set to gray with the white outline added to make the images easier to compare.

\subsection{Diffraction Spectrum Truncation}

High-order diffraction peaks are difficult to measure, both because their amplitudes tend to die off with increasing $n$, and because spanning large momentum transfer requires longer measurement times.  However, the resolution of the reconstruction will go like $ 2 \pi / Q_\mathrm{tot}$, where $Q_\mathrm{tot}$ is the total range in wavenumber transfer probed.  Thus, there is a cost-benefit to measuring higher diffraction orders versus taking more sample rotation steps.  However, tomographic reconstructions of the retrieved phase may also improve the spatial resolution of the reconstruction.  This is because lower diffraction orders in the measured diffraction spectra $\widetilde{S}(\beta)$ are a mixture of higher-order diffraction peaks.  For example, the $n=1$ diffraction peak in $\widetilde{S}(\beta)$, is a mixture of the $n=1$ Fourier coefficient of the underlying phase profile and the product of the $n=2$ and $n=-1$ coefficients, corresponding to the first-order and second-order term in the Born series for $e^{-i \phi(x)}$, respectively.  The optimal choice in whether larger diffraction orders or a higher density of sample rotations should be pursued will likely depend on the overall phase shift of the sample.  If samples with large phase shifts are measured, more information is contained in the amplitudes of the low diffraction orders as a function of sample rotation, as more terms in the Born series are relevant.  However, the samples considered here, both virtual and experimental, have overall amplitudes of less than $2 \pi$.  

\begin{figure}
\centering
\includegraphics[width=.95\linewidth]{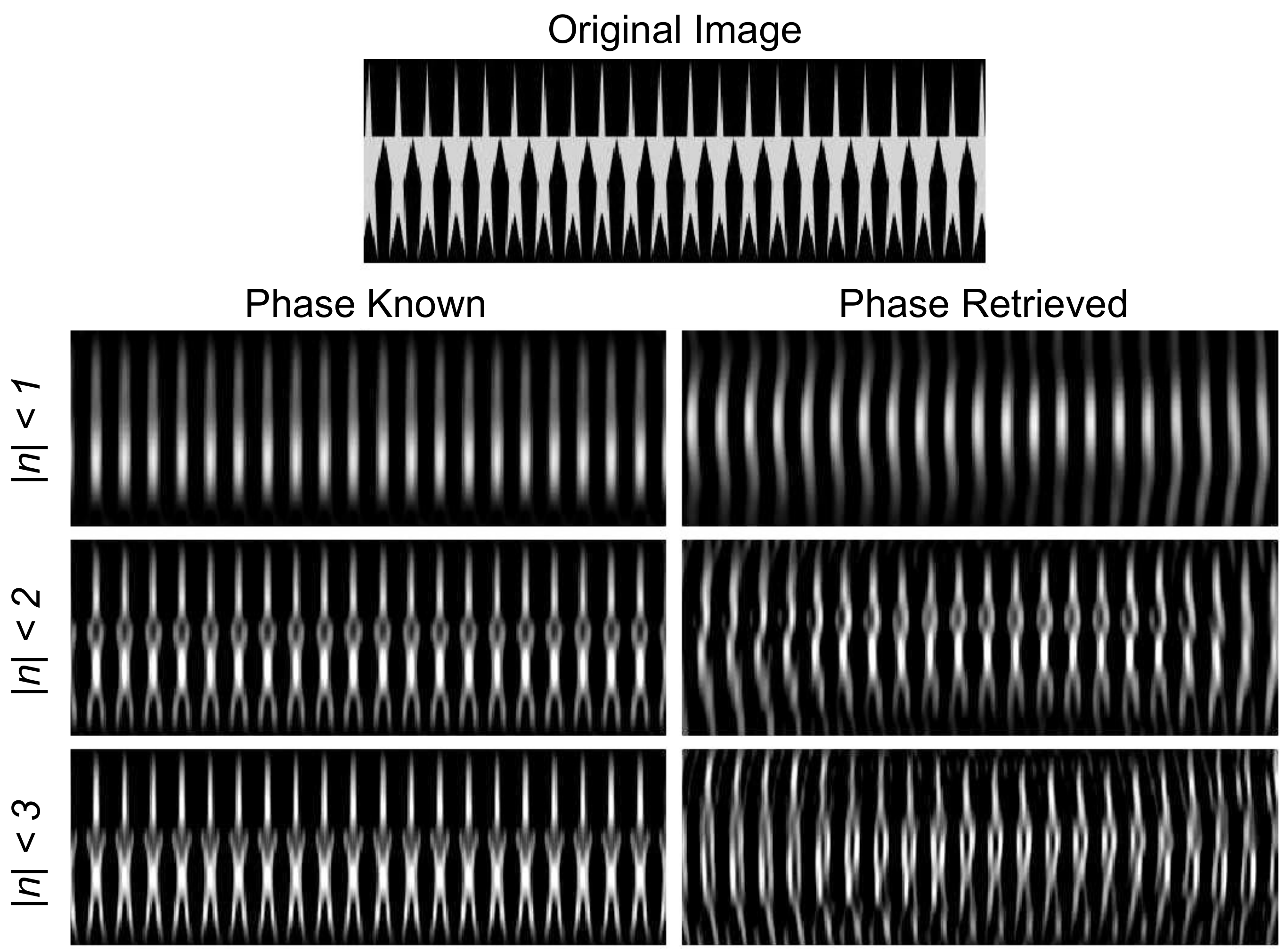}
\caption{Results of phase retrieval algorithm and tomographic reconstruction with the FFT of a Radon transformed image as inputs.  A comparison of reconstructions after truncating $\widetilde{S}(\beta)$ past $n^\mathrm{th}$ order (rows) with the phase of $\widetilde{S}(\beta)$ left intact (left column) and with the phase retrieved (right column).}
 \label{fig:phantom}
\end{figure}

To study the performance of the tomographic phase retrieval algorithm as well as how truncating $\widetilde{S}(\beta)$ affects the reconstructions, we simulated diffraction spectra from the image shown in Fig.~\ref{fig:phantom}.  Diffraction spectra were obtained from phase profiles of the Radon-transformed image from -19~degrees to 19~degrees in 2~degree steps.  The diffraction spectra were then set to zero after the first, second, or third-order diffraction peaks.  The real-space phase was then reconstructed with (1) the phase of the truncated $\widetilde{S}(\beta)$ left intact and (2) after passing $ \left | \widetilde{S}(\beta) \right |$ through the described phase retrieval algorithm.

The reconstructions of the original image are shown in Fig.~\ref{fig:phantom}.  The spatial resolution of the reconstructed images increases with increasing $|n|_\mathrm{max}$.  The reconstructions where the phase was retrieved provide good estimations of the reconstructions using the known phase.  However, some artifacts are evident for the phase-retrieved case.  There tends to be a region of high-fidelity, with distortions worsening further away horizontally.  The region that best estimates the original image is not necessarily centered, nor at the edge of the reconstruction because, as previously discussed in Section~\ref{sec:phaserec}, phase retrieval algorithms produce translationally-invariant solutions.  The distortions are of similar character for each level of diffraction order truncation.  Similar effects were seen in the reconstructions of the neutron phase-gratings.  The best region of the reconstruction was selected visually, as shown with the dashed line in Fig.~\ref{fig:Alg}.

\section{Conclusion}

We find that the neutron diffraction spectra of silicon phase-gratings as a function of grating rotation can be used to tomographically reconstruct the shape of the gratings.  These reconstructions rely on the periodic structure of the gratings, but nonetheless have a spatial resolution of around 300~nm, which is more than an order of magnitude smaller than other forms of neutron tomography.  In principle, even smaller structures may be probed, in which case the spatial resolution of the reconstruction is nominally $\sim 2 \pi / Q_\mathrm{tot}$, with $Q_\mathrm{tot} = {2 \pi \over \lambda} (\theta_\mathrm{max} - \theta_\mathrm{min})$, given by the rotational range of the double crystal diffractometer.  However, neutron scattering length densities for most materials are such that tens of micrometers of material are ordinarily required for a $2\pi$ phase amplitude given typical neutron wavelengths.  Thus it may be difficult to measure structures with amplitudes less than a hundred nanometers. 

An upper limit on the length scale to which this type of tomography is sensitive is set by the neutron coherence length of approximately 35~$\mu$m.  However, the addition of a PSD and possible combination of phase recovery with the USANS refractive signal \cite{treimer2003refraction} would allow for much larger reconstructions.  Combining phase recovery with other tomography signals is intended for future work.

Further optimization of the phase retrieval and reconstruction algorithms is likely possible.  For example, selection of the width of the Gaussian filter employed in the deconvolution step is related to the noise present in the measurement of the diffraction spectra.  Analysis of both real and simulated noisy data sets may elucidate how to best set this parameter.  Optimization of the cost-benefit between measuring larger ranges in the diffraction spectrum, versus taking a higher density of scans through $\beta$-space is also of interest.  This problem is likely dependent on the overall phase amplitude of the sample.  One may study how these and other changes to the algorithm impact the reconstruction of digital phantoms when the FFT power spectra of the Radon-transformed image are used as inputs for the reconstruction, similar to Fig.~\ref{fig:phantom}.  Despite the need for further improvements to the tomographic phase retrieval algorithm, the results presented here indicate that high quality tomographic reconstructions with sub-micrometer resolution of the retrieved phase are possible.

Two-dimensional phase retrieval and three-dimensional tomography could be achieved if data is taken as a function of more than one sample axis of rotation.  For a USANS setup, this would likely be a combination of $y$-axis and $z$-axis rotation in Fig.~\ref{fig:setup}, since only diffraction along the $x$-axis is measured.  For SANS data, where the diffraction spectra are inherently two-dimensional, only one axis of rotation is required for three-dimensional tomography. 

In the same way that neutron CT from other signals can be used to visualize materials in a way that is complementary to x-rays \cite{strobl2009advances}, phase-retrieved neutron tomography can provide unique information about a sample.  For example, a periodic structure buried in a matrix would produce a signal for neutrons, but may appear opaque to other forms of radiation, which could be especially beneficial for probing tissue scaffolds such as those described by \cite{dvir2011nanotechnological}.  In the case of phase-gratings, creating SEM micrographs entails cleaving the gratings, while our technique is both non-destructive and samples a much-larger area of the gratings.  Lithium-ion batteries can be imaged using neutrons \cite{siegel2011neutron}, and batteries with electrode layers that are too thin for traditional neutron imaging, such as those described in \cite{zhang2011three}, could benefit from this method.  If the algorithm is adapted to SANS data, in addition to measuring the spacing of oriented biological membranes \cite{nagy2011reversible}, SANS data could be used to probe the shape of the membranes.  The technique could also easily be extended to polarized beams to study magnetic samples.  For example, the depth profile of magnetic vortices could be probed \cite{eskildsen2009vortices,kawano2011gap,butch2018}.  Additionally, with the recent demonstration of measuring skyrmion lattices using SANS \cite{debeer2018realization}, it may be possible to generate three-dimensional renderings of skyrmions and provide important insights into their shape \cite{gilbert2018}.  Furthermore, any other sample with periodic structures that have a macroscopic ordering and can produce diffraction peaks in SANS or USANS data are candidates for phase-recovered neutron tomography.  For an unpolarized test case, we have successfully analyzed the shape of neutron phase-grating walls, confirming the tomographic reconstructions with SEM micrographs.

\section{Acknowledgments}

The authors would like to thank Markus Bleuel, Nicholas Butch, Brian Kirby, and Dustin Gilbert for discussing possible applications of phase-recovered neutron CT with us.  This work was supported by the U.S. Department of Commerce, the NIST Radiation and Physics Division, the Director's office of NIST, the NIST Center for Neutron Research, the NIST Quantum Information Program, the US Department of Energy under Grant No. DE-FG02-97ER41042, and National Science Foundation Grant No. PHY-1307426. This work was also supported by the Canadian Excellence Research Chairs (CERC) program, the Canada  First  Research  Excellence  Fund  (CFREF), the Natural Sciences and Engineering Research Council of Canada (NSERC) Discovery program, and the Collaborative Research and Training Experience (CREATE) program.

\bibliography{lib}

\end{document}